\documentclass[aps,prl,twocolumn,superscriptaddress,showpacs]{revtex4-1}

\usepackage{graphicx}
\usepackage{tabularx}

\begin{document}


\title{ 
Observation of Reactor Electron Antineutrino Disappearance in the RENO Experiment
}

%

\affiliation{Department of Physics, Chonbuk National University, Jeonju 561-756, Korea            }
\affiliation{Department of Physics, Chonnam National University, Gwangju, 500-757, Korea          }
\affiliation{Department of Physics, Chung Ang University, Seoul 156-756, Korea                    }
\affiliation{Department of Radiology, Dongshin University, Naju, 520-714, Korea                     }
\affiliation{Department of Physics, Gyeongsang National University, Jinju, 660-701, Korea         }
\affiliation{Department of Physics, Kyungpook National University, Daegu, 702-701, Korea          }
\affiliation{Department of Physics, Pusan National University, Busan, 609-735, Korea              }
\affiliation{Department of Physics, Sejong University, Seoul 143-747, Korea                       }
\affiliation{Division of General Education, Seokyeong University, Seoul, 136-704, Korea             }
\affiliation{Department of Physics and Astronomy, Seoul National University, Seoul 151-742, Korea }
\affiliation{Department of Fire Safety, Seoyeong University, Gwangju, 500-742, Korea              }
\affiliation{Department of Physics, Sungkyunkwan University, Suwon, 440-746, Korea                }

\author{J. K. Ahn}
\affiliation{Department of Physics, Pusan National University, Busan, 609-735, Korea              }
 \author{S. Chebotaryov}
\affiliation{Department of Physics, Kyungpook National University, Daegu, 702-701, Korea          }
\author{J. H. Choi}
\affiliation{Department of Radiology, Dongshin University, Naju, 520-714, Korea                     }
\author{S. Choi}
\affiliation{Department of Physics and Astronomy, Seoul National University, Seoul 151-742, Korea }
\author{W. Choi}
\affiliation{Department of Physics and Astronomy, Seoul National University, Seoul 151-742, Korea }
\author{Y. Choi}
\affiliation{Department of Physics, Sungkyunkwan University, Suwon, 440-746, Korea                }
\author{H. I. Jang}
\affiliation{Department of Fire Safety, Seoyeong University, Gwangju, 500-742, Korea              }
\author{J. S. Jang}
\affiliation{Department of Physics, Chonnam National University, Gwangju, 500-757, Korea          }
\author{E. J. Jeon}
\affiliation{Department of Physics, Sejong University, Seoul 143-747, Korea                       }
\author{I. S. Jeong}
\affiliation{Department of Physics, Chonnam National University, Gwangju, 500-757, Korea          }
\author{K. K. Joo}
\affiliation{Department of Physics, Chonnam National University, Gwangju, 500-757, Korea          }
\author{B. R. Kim}
\affiliation{Department of Physics, Chonnam National University, Gwangju, 500-757, Korea          }
\author{B. C. Kim}
\affiliation{Department of Physics, Chonnam National University, Gwangju, 500-757, Korea          }
\author{H. S. Kim}
\affiliation{Department of Physics, Chonbuk National University, Jeonju 561-756, Korea            }
\author{J. Y. Kim}
\affiliation{Department of Physics, Chonnam National University, Gwangju, 500-757, Korea          }
\author{S. B. Kim}
\affiliation{Department of Physics and Astronomy, Seoul National University, Seoul 151-742, Korea }
 \author{S. H. Kim}
\affiliation{Department of Physics, Pusan National University, Busan, 609-735, Korea              }
 \author{S. Y. Kim}
\affiliation{Department of Physics, Pusan National University, Busan, 609-735, Korea              }
\author{W. Kim}
\affiliation{Department of Physics, Kyungpook National University, Daegu, 702-701, Korea          }
\author{Y. D. Kim}
\affiliation{Department of Physics, Sejong University, Seoul 143-747, Korea                       }
\author{J. Lee}
\affiliation{Department of Physics and Astronomy, Seoul National University, Seoul 151-742, Korea }
\author{J. K. Lee}
\affiliation{Department of Physics, Pusan National University, Busan, 609-735, Korea              }
 \author{I. T. Lim}
\affiliation{Department of Physics, Chonnam National University, Gwangju, 500-757, Korea          }
\author{K. J. Ma}
\affiliation{Department of Physics, Sejong University, Seoul 143-747, Korea                       }
\author{M. Y. Pac}
\affiliation{Department of Radiology, Dongshin University, Naju, 520-714, Korea                     }
\author{I. G. Park}
\affiliation{Department of Physics, Gyeongsang National University, Jinju, 660-701, Korea         }
\author{J. S. Park}
\affiliation{Department of Physics and Astronomy, Seoul National University, Seoul 151-742, Korea }
 \author{K. S. Park}
\affiliation{Division of General Education, Seokyeong University, Seoul, 136-704, Korea             }
\author{J. W. Shin}
\affiliation{Department of Physics and Astronomy, Seoul National University, Seoul 151-742, Korea }
\author{K. Siyeon}
\affiliation{Department of Physics, Chung Ang University, Seoul 156-756, Korea                    }
\author{B. S. Yang}
\affiliation{Department of Physics and Astronomy, Seoul National University, Seoul 151-742, Korea }
\author{I. S. Yeo}
\affiliation{Department of Physics, Chonnam National University, Gwangju, 500-757, Korea          }
 \author{S. H. Yi}
\affiliation{Department of Physics, Sungkyunkwan University, Suwon, 440-746, Korea                }
\author{I. Yu}
\affiliation{Department of Physics, Sungkyunkwan University, Suwon, 440-746, Korea                }

\collaboration{RENO Collaboration}

%
%

\begin{abstract}

The RENO experiment has observed the disappearance of reactor electron antineutrinos, consistent with neutrino oscillations, with a significance of 4.9 standard deviations. Antineutrinos from six 2.8 GW$_{th}$ reactors at the Yonggwang Nuclear Power Plant in Korea, are detected by two identical detectors located at 294 m and 1383 m, respectively, from the reactor array center. In the 229 day data-taking period between 11 August 2011 and 26 March 2012, the far (near) detector observed 17102 (154088) electron antineutrino candidate events with a background fraction of 5.5\% (2.7\%). The ratio of observed to expected numbers of antineutrinos in the far detector is $0.920 \pm 0.009({\rm stat.}) \pm 0.014({\rm syst.})$. From this deficit, we determine $\sin^2 2 \theta_{13} = 0.113 \pm 0.013({\rm stat.}) \pm 0.019({\rm syst.})$ based on a rate-only analysis.
\end{abstract}


\maketitle

%

We report a definitive measurement of the neutrino oscillation mixing angle, $\theta_{13}$, based on the disappearance of electron antineutrinos emitted from six reactors. Of the three mixing angles in the Pontecorvo-Maki-Nakagawa-Sakata matrix \cite{ref_1, ref_2}, $\theta_{13}$ remains the most poorly known. Previous attempts at measuring $\theta_{13}$ via neutrino oscillations have obtained only upper limits [3$-$9]; the CHOOZ \cite{ref_3} and MINOS \cite{ref_5} experiments set the most stringent limits: $\sin^2 2\theta_{13} < 0.15$ (90\% C.L.). Recently, indications of a non-zero $\theta_{13}$ value have been reported by two accelerator appearance experiments, T2K \cite{ref_10} and MINOS \cite{ref_11}, and by the Double Chooz reactor disappearance experiment \cite{ref_12}. Global analyses of all available neutrino oscillation data have indicated central values of $\sin^2 2\theta_{13}$ that are between 0.05 and 0.1 (see e.g. \cite{ref_13, ref_14}). During the preparation of this paper, the Daya Bay experiment reported the measurement of a non-zero value for $\theta_{13}$ \cite{ref_15}.

Reactor experiments with a baseline distance of $\sim$1 km can neglect the disappearance of $\bar{\nu}_e$ driven by $\theta_{12}$ and $\Delta m_{21}^2$ and, thus, unambiguously determine the mixing angle $\theta_{13}$ based on the survival probability of electron antineutrinos, 
\begin{equation}
 P_{survival} \approx 1- \sin^2 2 \theta_{13} \sin^2(1.267 \Delta m_{31}^2 L/E),           
\end{equation}
where $E$ is the energy of antineutrinos in MeV, and $L$ is the baseline distance in meters between the reactor and detector. The well measured value of $\Delta m_{32}^2 = (2.32_{-0.08}^{+0.12})\times 10^{-3}$ eV$^2$ \cite{ref_16} can be substituted for $\Delta m_{31}^2$ in Eq. (1).

The detection methods and setup of the RENO experiment are discussed in detail elsewhere \cite{ref_17}. In this Letter, only the components relevant to this measurement are reviewed. Two identical antineutrino detectors are located at 294 m and 1383 m, respectively, from the center of reactor array to allow a relative measurement from a comparison of the measured neutrino rates. The measured far-to-near ratio of antineutrino fluxes can considerably reduce systematic errors coming from uncertainties in the reactor neutrino flux, target mass, and detection efficiency. The relative measurement is independent of correlated uncertainties and helps minimize uncorrelated reactor uncertainties. The far (near) detector is under a 450 (120) meters of water equivalent rock overburden.

Six pressurized water reactors, each with maximum thermal output of 2.8 GW$_{th}$ (reactors 3, 4, 5, and 6) or 2.66 GW$_{th}$ (reactors 1 and 2), are situated in a line with roughly equal spacings and span a total distance of $\sim$1.3 km. The positions of the two detectors and the six reactors were surveyed with GPS and total station to determine the baseline distances between the detectors and reactors to accuracies better than 10 cm. The reactor-flux weighted baseline is 408.56 m for the near detector, and 1443.99 m for the far detector. 

The reactor $\bar{\nu}_e$ is detected via the inverse beta decay (IBD) reaction, $\bar{\nu}_e + p \rightarrow e^+  + n$. Detectors based on hydrocarbon liquid scintillator (LS) provide free protons as a target. The coincidence of a prompt positron signal and a delayed signal from neutron capture by Gadolinium (Gd) provides the distinctive IBD signature.

\begin{figure}[hbt]
\begin{center}
\includegraphics[width=0.48\textwidth]{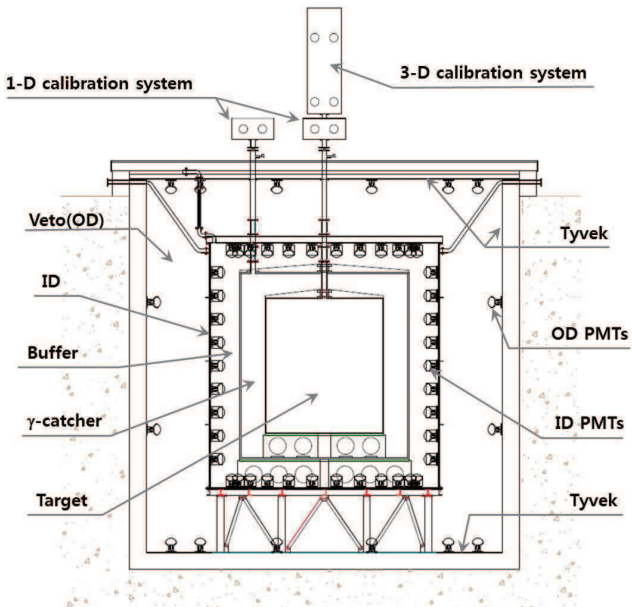}
\caption{A schematic view of a RENO detector. The near and far detectors are identical.}
\label{fig:RENO-detector}
\end{center}
\end{figure}
Each RENO detector (Fig. \ref{fig:RENO-detector}) consists of a main inner detector (ID) and an outer veto detector (OD). The main detector is contained in a cylindrical stainless steel vessel that houses two nested cylindrical acrylic vessels\cite{ref_18}. The innermost acrylic vessel holds the 18.6 m$^3$ (16 t) $\sim$0.1\% Gd-doped LS as a neutrino target. It is surrounded by a $\gamma$-catcher region with a 60 cm thick layer of Gd-unloaded LS inside an outer acrylic vessel. Outside the $\gamma$-catcher is a 70 cm thick buffer region filled with 65 tons of mineral oil. Light signals emitted from particles interacting in ID are detected by 354 10-inch Hamamatsu R7081 photomultiplier tubes (PMTs) that are mounted on the inner wall of the stainless steel container. The 1.5 m thick region of the OD that is external to the ID is filled with highly purified water. The OD is equipped with 67 10-inch R7081 PMTs mounted on the wall of the veto vessel.
The LS is developed and produced as a mixture of linear alkyl benzene (LAB), PPO, and bis-MSB. A Gd-carboxylate complex using TMHA was developed for the best Gd loading efficiency into LS and its long term stability \cite{ref_19}. 

Event triggers are formed by the number of PMTs with signals above a $\sim$0.3 photoelectron (p.e.) threshold (NHIT). An event is triggered and recorded for an IBD candidate if the ID NHIT is larger than 90, corresponding to 0.5$\sim$0.6 MeV and well below the 1.02 MeV minimum energy of an IBD positron signal. 
The ID trigger provides no loss of IBD candidates.

The detectors are calibrated using radioactive sources and cosmic-ray induced background event samples. Radioisotopes of $^{137}$Cs, $^{68}$Ge, $^{60}$Co, and $^{252}$Cf are periodically deployed in the target and $\gamma$-catcher by a step motorized pulley system in a glove box. The detectors' energy response stability is continuously monitored using cosmic-ray produced neutron captures on H and Gd.

The event energy is determined from the total p.e.-charge ($Q_{tot}$) that is collected by the PMTs, corrected for gain variation. The energy calibration constant of ~250 p.e. per MeV is determined from the peak energies of various radioactive sources deployed at the center of the target. The obtained energy resolution is 
$(5.9/ \sqrt{E({\rm MeV})}+1.1)\%$, common for both detectors.

In this analysis, an IBD event requires a delayed signal from a neutron capture on Gd and, thus, the fiducial volume naturally becomes the entire target vessel region without any vertex position cuts. There is some spill-in of IBD events that occur outside the target and produce a neutron capture on Gd in the target, which enhances the detection efficiency.

The following criteria are applied to select IBD candidate events: (1) $Q_{max}/Q_{tot} <$ 0.03 where $Q_{max}$ is the maximum charge of a PMT, to eliminate PMT flasher events and external $\gamma$-ray events; (2) a cut rejecting events that occur within a 1 ms window following a cosmic muon traversing the ID with an energy deposit ($E_{\mu}$) that is larger than 70 MeV, or with $E_{\mu}$ between 20 MeV and 70 MeV for OD NHIT $>$ 50; 
(3) events are rejected if they are within a 10 ms window following a cosmic muon  traversing the ID if $E_{\mu}$ is larger than 1.5 GeV; (4) 0.7 MeV $< E_p <$ 12.0 MeV; 
(5) 6.0 MeV $< E_d <$ 12.0 MeV where $E_p$ ($E_d$) is the energy of the prompt (delayed) event; (6) 2 $\mu$s $< \Delta t_{e^+ n} < $ 100 $\mu$s where $\Delta t_{e^+ n}$ is the time difference between the prompt and delayed signals; (7) a multiplicity requirement rejecting correlated coincidence pairs if they are accompanied by any preceding ID or OD trigger within a 100 $\mu$s window before their prompt candidate.

Applying the IBD selection criteria yields 17102 (154088) candidate events or 77.02$\pm$0.59 (800.8$\pm$2.0) events/day for a live time  of 222.06 (192.42) days in the far (near) detector. In the final data samples, some uncorrelated (accidentals) and correlated (fast neutrons from outside of ID, stopping muon followers, and $\beta$-n emitters from $^9$Li/$^8$He) background events survive the selection requirements. 
\begin{table}[hbt]
 \caption{Event rates of the observed candidates and the estimated background.}
 \begin{center}
 \begin{tabular*}{0.48\textwidth}{@{\extracolsep{\fill}} l r r }
 
  \hline \hline
     Detector              & Near       &  Far      \\
  \hline
 Selected events          &  154088     &   17102   \\
  
Total background rate (per day) &   21.75$\pm$5.93 &  4.24$\pm$0.75    \\
IBD rate after background       &   779.05$\pm$6.26 &  72.78$\pm$0.95   \\
~~~~~~~~~~~~~~subtraction (per day) &            &           \\
  \hline
DAQ Live time (days)       &   192.42      &    222.06    \\
Detection efficiency ($\epsilon$) & 0.647$\pm$0.014 & 0.745$\pm$0.014  \\
    \hline
  Accidental rate (per day)   &      4.30$\pm$0.06  &   0.68$\pm$0.03  \\
  $^9$Li/$^8$He rate (per day)&     12.45$\pm$5.93  &   2.59$\pm$0.75   \\
  Fast neutron rate (per day) &      5.00$\pm$0.13  &   0.97$\pm$0.06   \\
  \hline \hline
  \end{tabular*}
 \end{center}

 \label{tab:Event_rate}

 \end{table}

The uncorrelated background is due to accidental coincidences from the random association of a prompt-like event due to radioactivity and a delayed-like neutron capture. The remaining rate in the final sample is estimated by measuring the rates of prompt- and delayed-like events after applying all the selection criteria other than (6), and calculating the probability of random association in the 
$\Delta t $ window for IBD selection, leading to 4.30$\pm$0.06 (near) or 0.68$\pm$0.03 (far) events per day.

The $^9$Li/$^8$He $\beta$-n emitters are mostly produced by energetic muons because their production cross sections in carbon increase with muon energy [20$-$22]. The background rate is estimated from a sample prepared by a delayed coincidence between an energetic ($E_{\mu} >$ 0.5 GeV) muon and the following IBD-like pair of events. The $^9$Li/$^8$He $\beta$-n background rate in the final sample is obtained as 12.45$\pm$5.93 (near) or 2.59$\pm$0.75 (far) events per day from a fit to the delay time distribution with an observed mean decay time of $\sim$250 ms.

An energetic neutron entering the ID can interact in the target to produce a recoil proton before being captured on Gd. Fast neutrons are produced by cosmic muons traversing the surrounding rock and the detector. The background rate is estimated by extrapolating the energy spectral shape of events with 12 MeV $< E_p <$ 30 MeV, to the IBD signal region, assuming a flat spectrum of the fast neutron background. The estimated fast neutron background is 5.00$\pm$0.13 (near) or 0.97$\pm$0.06 (far) events per day. The total background rate is estimated to be 21.75$\pm$5.93 (near) or 4.24$\pm$0.75 (far) events per day and summarized in Table \ref{tab:Event_rate}. 

Both the prompt energy and flasher requirements are almost fully (99.8\%) efficient. The fraction of neutron captures on Gd is evaluated to be (85.5$\pm$0.7)\% using MC and $^{252}$ Cf source data. The $\Delta t_{e^+ n}$ requirement efficiency is determined to be (92.1$\pm$0.5)\% from MC and data. The fraction of neutron captures on Gd that satisfy the 6.0 MeV threshold requirement is (95.2$\pm$0.5)\%. The overall efficiency for finding a delayed signal as an IBD candidate pair is (74.9$\pm$1.0)\%. The spill-in IBD events result in a net increase in the detection efficiency of 2.2\%. The common detection efficiency is estimated to be (76.5$\pm$1.4)\% using a Monte Carlo simulation (MC) and data. 

The fractional losses of IBD events due to the muon veto are determined to be (11.30$\pm$0.04)\% (near) or (1.36$\pm$0.02)\% (far), by summing the time spent in vetoing events after muons. The fractional losses of IBD events due to the multiplicity cut is calculated to be (4.61$\pm$0.04)\% (near) or (1.22$\pm$0.07)\% (far), based on the ID trigger rate and the veto window from an IBD prompt candidate. The efficiencies for detecting IBD events are found to be (64.7$\pm$1.4)\% (near) and (74.5$\pm$1.4)\% (far).

The absolute uncertainties of the efficiencies are correlated between the two detectors. Only differences between the two
identical detectors are taken as uncorrelated uncertainties. The systematic uncertainties are summarized in Table II.
\begin{table}[hb]
 \caption{Systematic uncertainties in the reactor neutrino detection.}
 \begin{center}
 \begin{tabular*}{0.48\textwidth}{@{\extracolsep{\fill}} l r r }

    \hline \hline
    \multicolumn{3}{ c }{\bf Reactor}                       \\
  \hline
                      &   Uncorrelated  &    Correlated   \\
  \hline
   Thermal power      &   0.5\%          &      $-$       \\
   Fission fraction   &   0.7\%          &      $-$       \\
   Fission reaction cross section &  $-$ &  1.9\%         \\
   Reference energy spectra       &   $-$  &  0.5\%         \\
   Energy per fission &     $-$            &  0.2\%         \\
  \hline
   Combined            &     0.9\%         &  2.0\%         \\ 
  \hline \hline
    \multicolumn{3}{ c }{\bf Detection}                      \\
  \hline
                      &   Uncorrelated     &    Correlated   \\
  \hline
   IBD cross section  &       $-$        &      0.2\%     \\
   Target protons     &     0.1\%        &      0.5\%     \\  
   Prompt energy cut  &     0.01\%       &      0.1\%    \\
   Flasher cut        &     0.01\%       &      0.1\%     \\
   Gd capture ratio   &     0.1\%        &      0.7\%    \\
   Delayed energy cut  &    0.1\%      &      0.5\%     \\
   Time coincidence cut &   0.01\%      &        0.5\%     \\
   Spill-in             &    0.03\%      &        1.0\%     \\ 
  Muon veto cut       &     0.02\%        &       0.02\%      \\
  Multiplicity cut    &     0.04\%        &       0.06\%       \\                     
   \hline
   Combined (total)    &      0.2\%       &     1.5\%        \\
  \hline \hline
  \end{tabular*}
 \end{center}
 
 \label{tab:Systematic}

 \end{table}

Uncorrelated relative uncertainties are estimated by comparing the two detectors. The IBD differential cross section is taken from Ref. \cite{ref_23}. The total number of free protons in the target is $1.189 \times 10^{30}$ with an uncertainty of 0.5\%, determined from measurements of the LS weight and composition. The relative energy scale difference between the detectors is determined to be 0.2\% from comparison of the peak energy values for several radioactive calibration sources, IBD delayed events, and cosmic muon induced spallation-neutron captures on H and Gd. The energy scale difference is found to correspond to a relative uncertainty in the efficiency of the delayed energy of 0.1\% using data. The Gd-LS was commonly produced and then divided equally and filled into the two detectors to ensure that the Gd concentration and the target protons of the near and far detectors are identical. This procedure for filling the targets results in a difference in the number of the target protons that is less than 0.1\%. The difference in the measured neutron capture time between the detectors is less than 0.2 $\mu$s, corresponding to Gd concentration differences of less than 0.1\%. The relative uncertainty of Gd capture ratio is less than 0.1\% accordingly.
The remaining relative uncertainties are close to 0.01\%, and the combined uncertainty common to the both detectors is 0.2\%. A more detailed discussion on the systematic uncertainties will be presented in a future publication. 

The antineutrino flux depends on thermal power, fission fractions of the four isotopes, energy released per fission, and fission and capture cross-sections. The uncertainty associated with the thermal power, provided by the power plant, is 0.5\% per core and fully correlated among the reactors \cite{ref_24}. The relative fission contributions of the four main isotopes are evaluated for the fuel cycle with 4$\sim$10\% uncertainties, using the Westinghouse ANC reactor simulation code \cite{ref_25}. The uncertainties of the fission fraction simulation contribute 0.7\% of the $\bar{\nu}_e$ yield per core to the uncorrelated uncertainty. The associated antineutrino flux is computed based on the $\bar{\nu}_e$ yield per fission \cite{ref_26} and the fission spectra [27$-$31], leading to a 1.9\% correlated uncertainty that has little effect on the $\theta_{13}$ determination. The thermal energy released per fission is given in Ref. \cite{ref_32}, and its uncertainty results in a 0.2\% correlated uncertainty. We assume a negligible contribution of the spent fuel to the uncorrelated uncertainty in this analysis. 

\begin{figure}[hbt]
\begin{center}
\includegraphics[width=0.47\textwidth]{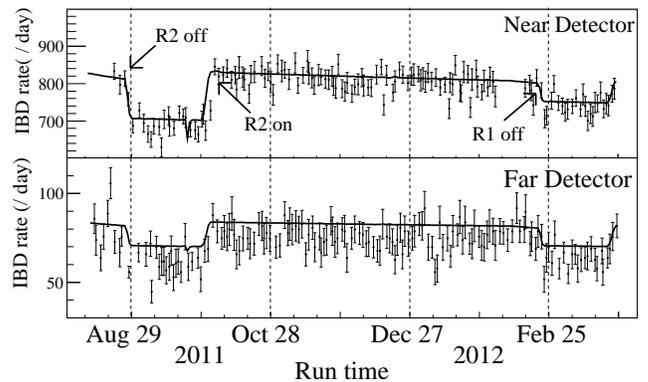}
\caption{Measured daily-average rates of reactor neutrinos after background subtraction in the near and far detectors as a function of running time. The solid curves are the predicted rates for no oscillation.}
\label{fig:daily-neutrino-rate}
\end{center}
\end{figure}

All reactors were mostly in steady operation at the full power during the data-taking period, except for reactor 2 (R2), which was off for the month of September 2011, and reactor 1 (R1), which was off from February 23 2012 for fuel replacement. Figure \ref{fig:daily-neutrino-rate} presents the measured daily rates of IBD candidates after background subtraction in the near and far detectors. The expected rates assuming no oscillation, obtained from the weighted fluxes by the thermal power and the fission fractions of each reactor and its baseline to each detector, are shown for comparison.

The ratio of measured to expected events in the far detector is
\begin{center}
$R$ = 0.920 $\pm$ 0.009(stat.) $\pm$ 0.014(syst.),
\end{center}
which indicates a clear deficit. To determine the value of $\sin^2 2\theta_{13}$ from the deficit, a $\chi^2$ with pull terms on the correlated systematic uncertainties \cite{ref_33} is used,
\begin{eqnarray}
 \chi^2  & = & \sum_{d=N, F} \frac{ \left[ N_{obs}^d + b_d
  -(1+a+\xi_d )\sum_{r=1}^{6}(1+f_r)N_{exp}^{d,r} \right]^2 } {N_{obs}^d }
 \nonumber       \\
&& + \sum_{d=N, F} \left( \frac{\xi_d^2}{{\sigma_d^{\xi}}^2} + \frac{b_d^2}{{\sigma_d^b}^2} \right) + \sum_{r=1}^{6} \left( \frac{f_r}{\sigma_r} \right)^2 ,  
\end{eqnarray}
where $d$ is an index denoting the near detector (N) or the far detector (F), $r$ corresponds to reactors 1 through 6, $N_{obs}^d$ is the number of observed IBD candidates in each detector after background subtraction, and $N_{exp}^{d, r}$ is the number of expected neutrino events, including detection efficiency, neutrino oscillations, and contribution from the $r$-th reactor to each detector determined from baseline distances and reactor fluxes. A global normalization $a$ is taken free and determined from the fit to the data. Then, $a$ is constrained by the normalization uncertainty of 2.5\%, coming from correlated uncertainties, to the value obtained from the fit. The uncorrelated reactor uncertainty is 0.9\% ($\sigma_r$), the uncorrelated detection uncertainty is 0.2\% ($\sigma_d^{\xi}$), as listed in Table \ref{tab:Systematic}, and $\sigma_d^b$ is the background uncertainty listed in Table \ref{tab:Event_rate}. $f_r$, $\xi_d$, and $b_d$ are corresponding pull parameters.

The best-fit value thus obtained is 
\begin{equation}
\sin^2 2\theta_{13} = 0.113 \pm 0.013(\rm stat.) \pm 0.019(\rm syst.),
\end{equation}
and excludes the no-oscillation hypothesis at the 4.9 standard deviation level.            

\begin{figure}[hbt]
\begin{center}
\includegraphics[width=0.47\textwidth]{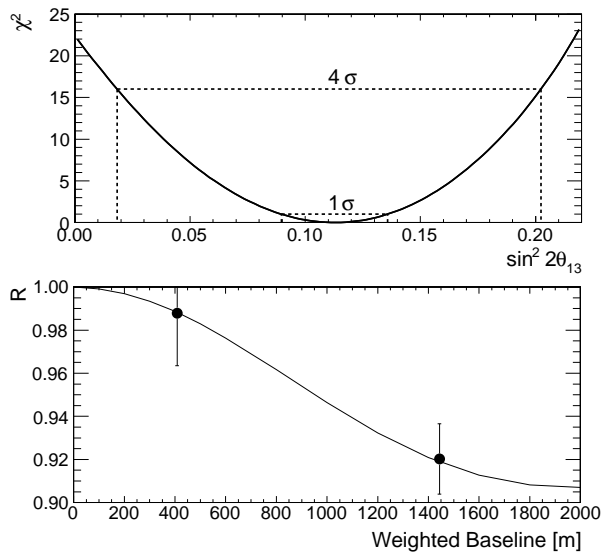}
\caption{The $\chi^2$ distribution as a function of $\sin^2 2\theta_{13}$. Bottom: Ratio of the measured reactor neutrino events relative to the expected with no oscillation. The curve represents the oscillation survival probability at the best fit, as a function of the flux-weighted baselines.}
\label{fig:chi-square}
\end{center}
\end{figure}

Figure \ref{fig:chi-square} shows the $\chi^2$ distribution as a function of $\sin^2 2\theta_{13}$, and the ratios of the measured reactor neutrino events, relative to what is expected without oscillation at both detectors.
We observe a clear deficit of 8.0\% for the far detector, and of 1.2\% for the near detector, concluding a definitive observation of reactor antineutrino disappearance consistent with neutrino oscillations. The survival probability due to neutrino oscillation at the best-fit value is given by the curve.

\begin{figure}[hbt]
\begin{center}
\includegraphics[width=0.47\textwidth]{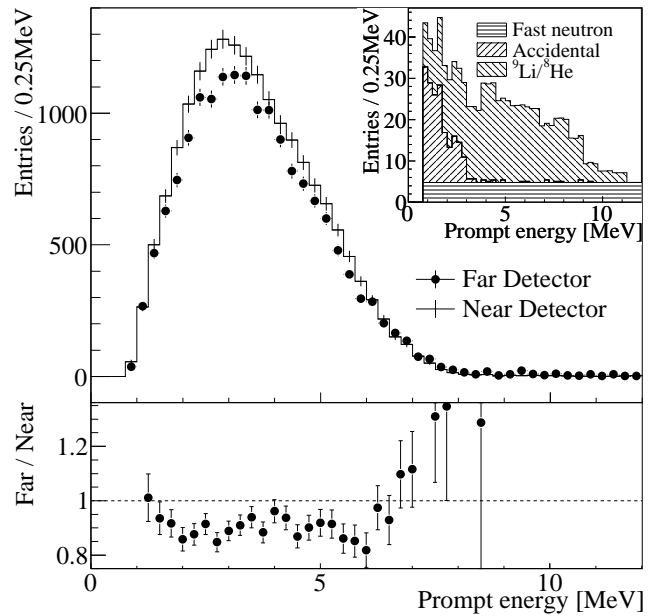}
\caption{Observed spectrum of the prompt signals in the far detector compared with the non-oscillation predictions from the measurements in the near detector. The backgrounds shown in the inset are subtracted for the far spectrum. The background fraction is 5.5\% (2.7\%) for far (near) detector. Errors are statistical uncertainties only. Bottom: The ratio of the measured spectrum of far detector to the non-oscillation prediction.}
\label{fig:prompt-spectra}
\end{center}
\end{figure}

The observed spectrum of IBD prompt signals in the far detector is compared to non-oscillation expectations based on measurements in the near detector in Fig. \ref{fig:prompt-spectra}.
The spectra of prompt signals are obtained after subtracting backgrounds  shown in the inset.
The disagreement of the spectra provides further evidence of neutrino oscillation. 

In summary, RENO has observed reactor antineutrinos using two identical detectors each with 16 tons of Gd-loaded liquid scintillator, and a 229 day exposure to six reactors with total thermal energy 16.5 GW$_{th}$. In the far detector, a clear deficit of 8.0\% is found by comparing a total of 17102 observed events with an expectation based on the near detector measurement assuming no oscillation. From this deficit, a rate-only analysis obtains $\sin^2 2 \theta_{13}$ = 0.113 $\pm$ 0.013(stat.) $\pm$ 0.019(syst.). The neutrino mixing angle $\theta_{13}$ is measured with a significance of 4.9 standard deviation.

The RENO experiment is supported by the Ministry of Education, Science and Technology of Korea and the Korea Neutrino Research Center selected as a Science Research Center by the National Research Foundation of Korea (NRF). Some of us have been supported by a fund from the BK21 of NRF.
We gratefully acknowledge the cooperation of the Yonggwang Nuclear Power Site and the Korea Hydro \& Nuclear Power Co., Ltd. (KHNP). 
We thank KISTI's providing computing and network resources through GSDC, and all the technical and administrative people who greatly helped in making this experiment possible.


\begin{thebibliography}{}




\bibitem{ref_1} B. Pontecorvo,  Zh. Eksp. Theo. Fiz. {\bf 34}, 247 (1957) [Sov. Phys. JETP {\bf 7}, 172 (1958)].
\bibitem{ref_2} Z. Maki, M. Nakagawa, and S. Sakata, Prog. Theor. Phys. {\bf 28}, 870 (1962).
\bibitem{ref_3} M. Apollonio {\it et al}. (Chooz Collaboration), Phys. Lett. {\bf B466}, 415 (1999); Eur. Phys. J. C {\bf 27}, 331 (2003).
\bibitem{ref_4} F. Boehm {\it et al}. (Palo Verde Collaboration), Phys. Rev. Lett. {\bf 84}, 3764 (2000).
\bibitem{ref_5} P. Adamson {\it et al}. (MINOS Collaboration), Phys. Rev. D {\bf 82}, 051102 (2010).
\bibitem{ref_6} S. Yamamoto {\it et al}. (K2K Collaboration), Phys. Rev. Lett. {\bf 96}, 181801 (2006).
\bibitem{ref_7} R. Wendell {\it et al}. (Super-Kamiokande Collaboration), Phys. Rev. D {\bf 81}, 092004 (2010).
\bibitem{ref_8} B. Aharmim {\it et al}. (SNO Collaboraiton), Phys. Rev. C {\bf 81}, 055504 (2010).
\bibitem{ref_9} A. Gando {\it et al}. (KamLAND Collaboration), Phys. Rev. D {\bf 83}, 052002 (2011).
\bibitem{ref_10} K. Abe {\it et al}. (T2K Collaboration), Phys. Rev. Lett. {\bf 107}, 041801 (2011).
\bibitem{ref_11} P. Adamson {\it et al}. (MINOS Collaboration), Phys. Rev. Lett. {\bf 107}, 181802 (2011).
\bibitem{ref_12} Y. Abe {\it et al}. (Double Chooz Collaboration), Phys. Rev. Lett. {\bf 108}, 131801 (2012).
\bibitem{ref_13} G. L. Fogli {\it et al}., Phys. Rev. D {\bf 84}, 053007 (2011).
\bibitem{ref_14} T. Schwetz {\it et al}., New J. Phys. {\bf 13}, 109401 (2011).
\bibitem{ref_15} F. P. An {\it et al}. (Daya Bay Collaboration) (2012), arXiv:hep-ex/1203.1669.
\bibitem{ref_16} P. Adamson {\it et al}. (MINOS Collaboration), Phys. Rev. Lett. {\bf 106}, 181801 (2011).
\bibitem{ref_17} J.K. Ahn, {\it et al}. (RENO Collaboration) (2010), arXiv:hep-ex/1003.1391.
\bibitem{ref_18} K. S. Park, {\it et al}., Construction and Properties of Acrylic Vessels in the RENO Detector, in preparation.
\bibitem{ref_19} J. S. Park, {\it et al}., Production and Optical Properties of Gd-loaded Liquid Scintillator for the RENO Neutrino Detector, in preparation.
\bibitem{ref_20} T. Hagner {\it et al}., Astropart. Phys. {\bf 14}, 33 (2000).
\bibitem{ref_21} D. R. Tilley {\it et al}., Nucl. Phys. {\bf A745}, 155 (2004).
\bibitem{ref_22} S. Abe {\it et al}. (KamLAND Collaboration), Phys. Rev. C {\bf 81}, 025807 (2010).
\bibitem{ref_23} P. Vogel and J. F. Beacom, Phys. Rev. D {\bf 60}, 053003 (1999).
\bibitem{ref_24} S. F. E. Tournu {\it et al}., EPRI 2001.1001470, Palo Alto, CA. (2001).
\bibitem{ref_25} ANC: A Westinghouse Advanced Nodal Computing Code, Westinghouse Report WCAP-10965-P-A(P) (1986).
\bibitem{ref_26} Y. Declais {\it et al}., Phys. Lett. {\bf B338}, 383 (1994).
\bibitem{ref_27} W. G. K. Schreckenbach, G. colvin and F. von Feilitzsch,  Phys. Lett. {\bf B160}, 325 (1985).
\bibitem{ref_28} F. von Feilitzsch and K. Schreckenbach, Phys. Lett. {\bf B118}, 162 (1982).
\bibitem{ref_29} A. A. Hahn {\it et al}., Phys. Lett. {\bf B218}, 365 (1989).
\bibitem{ref_30} T. Mueller {\it et al}., Phys. Rev. C {\bf 83}, 054615 (2011).
\bibitem{ref_31} P. Huber, Phys. Rev. C {\bf 84}, 024617 (2011) [Erratum-ibid, {\bf 85}, 029901(E) (2012)].
\bibitem{ref_32} V. Kopeikin {\it et al}., Phys. Atom. Nucl. {\bf 67}, 1892 (2004).
\bibitem{ref_33} D. Stump {\it et al}., Phys. Rev. D {\bf 65}, 014012 (Appendix B) (2001).


\end{thebibliography}
\end{document}